\documentclass[
    ,final            
  ]
  {aipproc}

\layoutstyle{8x11single}

\usepackage{psfrag}
\newcommand {\onehalf}{\frac{1}{2}}

\begin{document}

\title{Hadronic decay width from finite-volume energy spectrum in lattice QCD}


\author{Pietro Giudice}
{
  address={School of Mathematics, Trinity College, Dublin 2, Ireland}
}

\author{Michael J. Peardon}{
  address={School of Mathematics, Trinity College, Dublin 2, Ireland}
}

\begin{abstract}
The standard approach to determine the parameters of a resonance
is based on the study of the volume dependence of the energy spectrum.
In this work we study a non-linear sigma model coupled to a scalar
field in which a resonance emerges. Using an analysis method introduced
recently, based on the concept of probability distribution, 
it is possible to determine the mass and the width of the resonance. 
\end{abstract}

\maketitle


\section{Introduction}

Lattice simulations provide a non-perturbative framework for
calculating many low energy properties of $QCD$. Of these studies, 
the mass spectrum of the lightest hadrons is amongst the most important.
In a lattice calculation the (stable) hadron spectrum is determined
by numerical measurement of the energy eigenvalues of the $QCD$
Hamiltonian. Nonetheless, most hadronic states are resonances which
cannot be identified with a single energy level of the Hamiltonian:
this complicates their mass and the width determination.
A standard approach, introduced by 
L\"uscher~\cite{Luscher:1986pf,Luscher:1990ux,Luscher:1991cf}
(see also~\cite{Wiese:1988qy,DeGrand:1990ip} and for a generalization 
in moving frames~\cite{Rummukainen:1995vs}), to extract 
resonance parameters, starts by studying the volume dependence 
of the energy spectrum of the Hamiltonian when the system is confined 
in a finite box. There is a connection between the 
two-particle energy spectrum in finite volume and the elastic scattering 
phase shift in the infinite volume.

It was noted that where the phase shift in infinite volume 
quickly passes through $\pi/2$ an abrupt rearrangement of the finite volume
energy levels occurs. This peculiar behaviour, known as an avoided level 
crossing (ALC), is the signature of a resonance in the lattice data.
Therefore, it is possible to extract  the phase shift from lattice simulations
and, consequently, to determine the position (the mass value) and the width 
of the resonances.
The feasibility of this method has been demonstrated in four dimensional 
Ising model~\cite{Montvay:1987us} and for models in two
dimensions: in particular in the $O(3)$ non-linear 
$\sigma-$model~\cite{Luscher:1990ck} and in two coupled Ising 
models~\cite{Gattringer:1992yz}.
Subsequently, it was applied in three dimensions in $QED$ to study the 
meson-meson scattering phase shift~\cite{Fiebig:1994qi} and 
in four dimensional $O(4)$ $\phi^4$ theory~\cite{Gockeler:1994rx}.
Recently the method has been used to extract the $\rho$ resonance from 
lattice $QCD$~\cite{Gockeler:2008kc,Aoki:2007rd} and to calculate the S-wave 
pion-pion scattering length in the isospin $I = 2$ channel and the P-wave 
pion-pion scattering phase in the isospin $I = 1$ channel~\cite{Feng:2009ck}.

During the last years many steps have been taken to address the problem
of resonances on lattice; however, the approach described before
is limited. Many of the resonances in $QCD$ are characterized 
by a large width and so the rearrangement of the finite-volume 
energy levels does not provide a clear signature of the presence of a resonance:
the ALC is washed out. A possible solution to this problem
comes from a new method to analyze the finite volume energy 
spectrum~\cite{Bernard:2008ax}.
The idea behind this work is to test this method in the case of a simple 
field model which is an effective low-energy description of $QCD$.
Moreover, we can take advantage of the results in this model to 
quantitatively predict the precision of the results in the 
determination of parameters of resonances in $QCD$; 
we plan to deal with the glueball resonance.

\section{Theoretical background}

In lattice field theory we determine the mass of a stable particle from 
the exponential decay of suitable correlation functions in Euclidean space.
If $G(x,y)$ is the two-point Green function of a generic field $\phi$, 
we can consider the \emph{partial} Fourier transform
\begin{equation}
C(t,\vec{p})=\int\frac{d^3\vec{x}}{(2\pi)^3} e^{-i\vec{p}\vec{x}}G(\vec{x},t;
\vec{0},0)=
\int \frac{d\omega}{2 \pi} e^{i \omega t} \tilde{G}(\vec{p},\omega) \ ,
\end{equation}
where $\tilde{G}(\vec{p},\omega)$ is the Fourier transform of $G(x,y)$.
One then expects that at large times $t$
\begin{equation}
C(t,\vec{p}=0) \propto e^{-m_{\phi} t} \ ,
\end{equation}
where $m_{\phi}$ is the lightest mass which appears in the spectrum of 
the theory.
On the other hand, if we consider an unstable particle $\phi$ which
can decay in two particles $\pi$ ($\phi \rightarrow 2 \pi$) 
when $m_\phi > 2 m_\pi$, at large times $t$ we have
\begin{equation}
C(t,\vec{p}=0) \propto e^{-2 m_\pi t} \ ,
\end{equation}
and so we have no information about the mass of $m_{\phi}$.

Theoretically, we can characterize a resonance studing the $S$-matrix
theory; let us consider the element $S_l(E)$ ($l$ is the angular momentum
and $E$ is the energy, considered as a complex variable) then we can define
a resonance as the pole $\tilde{E}$ of $S_l(E)$, in the complex energy plane, 
when Im$(\tilde{E}) < 0$ and Re$(\tilde{E}) \neq 0$.
In fact, if Im$(\tilde{E})$ is small it is possible to show 
that the presence of the pole appears as a \emph{peak} in the total cross 
section, for the corresponding $l$-wave contribution:
\begin{equation}
\sigma_l(E)  \propto \frac{1}{E} \frac{\Gamma^2/4}{(E-M)^2+\Gamma^2/4}
\end{equation}
where $M \propto \mbox{Re}(\tilde{E})$ and  
$\Gamma \propto \mbox{Im}(\tilde{E})$.
As a consequence, a resonance is characterized not only by a mass $M$ 
but also by a width $\Gamma$; therefore, we cannot identify a resonance 
with an isolated energy level. 
This has to be compared with the fact that the energy 
spectrum of the Euclidean theory is always real, i.e. we can only study 
isolated levels.  It is clear that it is not possible to get the parameters of 
a resonance using the standard tools of the lattice approach.

As explained in the introduction, if a resonance is present
in a theory, it manifests itself as a rearrangement of the energy levels
related to a mixing between the unstable particle $\phi$ and the two 
stable particles $\pi$; taking advantage of this effect, i.e. using 
the  L\"uscher approach, it is possible to obtain the resonance parameters.

\subsection{Interacting particles in a box}
Let us consider a system of two identical non-interacting bosons of mass
$m_\pi$ in a box of volume $V=\prod_{i=1}^3 L_i$. The two-particle states
are characterized by a total energy $E$, in the rest frame of the box, and
by a relative momentum $\vec{p}$ related by:
\begin{equation}
E=2 \sqrt{m_\pi^2+|\vec{p}|^2}.
\label{eqW}
\end{equation}
The quantized momentum $p_i$ are given by $p_i=\frac{2 \pi}{L_i} n_i$, 
where $n_i \in Z$. 
Note that Eq.~\eqref{eqW} is only an approximation on the lattice because it
does not take in account space-time discretization artefacts.

\begin{figure}[ht]
  \footnotesize
  \psfrag{L}{L}
  \psfrag{E}{E}
  \psfrag{PHI}{$m_\phi$}
  \includegraphics[width=0.48\textwidth]
{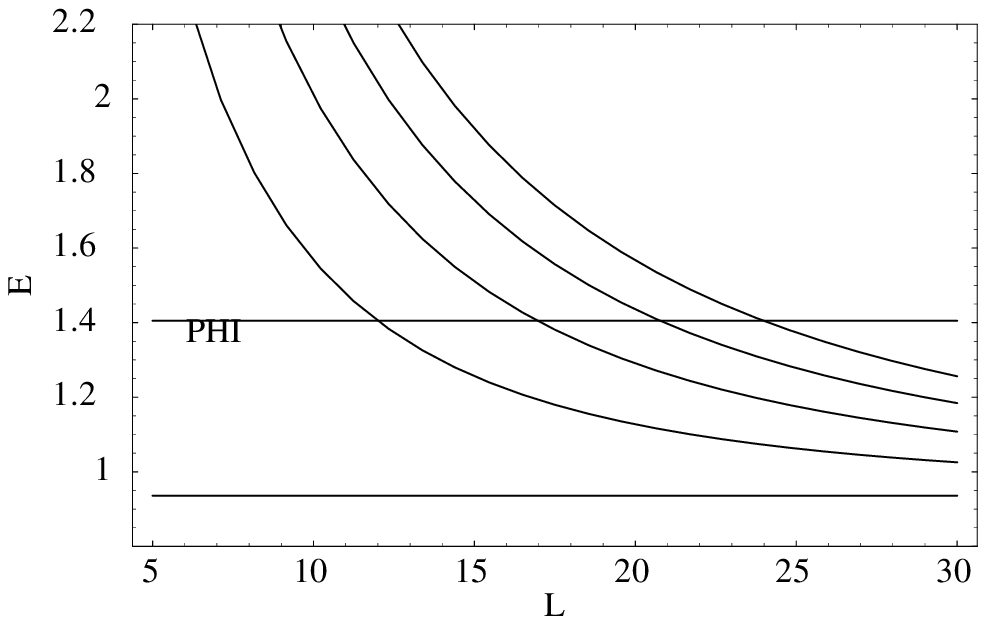}
\hspace{8mm}
  \includegraphics[width=0.48\textwidth]
{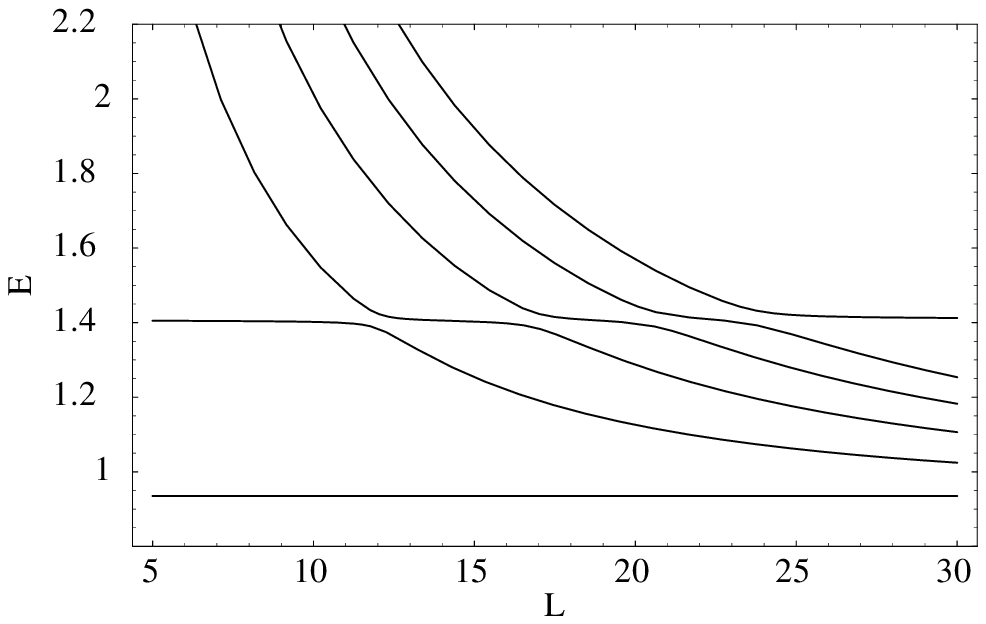}
\caption{(Left) The spectrum of a system of two non-interacting particles 
of mass $m_\pi=0.468$; the horizontal line describes the particle 
$\phi$ at rest of mass $m_\phi=1.405$. 
With these parameters the intersection between $\phi$ and the two-particle 
level $n^2=1$ is set at $L=12$. (Right) Avoided level crossings where on the 
Left there were intersections between $\phi$ and $2\pi$.}
\label{spectrum}
\end{figure}

We can study the spectrum of this system using Eq.~\eqref{eqW}; in particular,
we can determine it for different values of $L_i$ and for different values
of $\vec{n}$ ($\vec{n}=(n_1,n_2,n_3)$). In a cubic box, if 
$n^2=\sum_{i=1}^3 n_i^2$ is fixed, we have degenerate energy levels for 
different values of $n_i$. In Figure~\ref{spectrum} (Left) 
we show the five lowest levels as function of $L$ (cubic box).
As proposed in~\cite{Li:2007ey} can be an advantage to study the scattering 
using an asymmetric box; in fact, this allows to access more non-degenerate 
low-momentum modes for a given volume.

Because we are are interesting in resonant scattering, let us introduce 
a new particle in the system with mass $m_\phi$ related to the physical mass
of $\pi$ by $2 m_\pi < m_\phi < 4 m_\pi$; at the moment, it is not 
interacting with the particles $\pi$. In Figure~\ref{spectrum} (Left)
the $\phi$ energy level is the horizontal line; 
it intersects the two-particle levels at various system sizes $L$.
If we introduce a three point interaction between $\phi$ and $\pi$,
in a Minkowski space and in an infinite volume, we can say the particle $\phi$
becames unstable and it decays into two $\pi$-particles. In Euclidean space 
and in a
finite volume we can only say that, because of the interaction, the energy 
eigenstates are a mixture of $\phi$ and $2\pi$. This is manifested as
ALCs in the energy levels as shown in Figure~\ref{spectrum} (Right).

When the interaction is turned on Eq.~\eqref{eqW} is still valid, but the 
momentum quantization condition is replaced with the L\"uscher formula 
which connects the momentum $p=|\vec{p}|$ with the scattering phase shift 
$\delta_l$:
\begin{equation}
\delta_l(p)=-\phi(q) \bmod \pi, \quad q=\frac{p L}{2 \pi},
\label{luschereq}
\end{equation}
where $\phi(q)$ is determined when $l=0$ by
\begin{equation}
\tan{\phi(q)}=-{\frac{\pi^{3/2} q}{Z_{00}(1;q^2)}}, \quad \phi(0)=0.
\end{equation}
The function $Z_{00}(s;q^2)$ is the generalized zeta function defined 
in~\cite{Luscher:1991cf}.
The determination of the resonance parameters follows immediately.
Using Monte Carlo simulations we can determine the two-particle energy 
spectrum (a plot looking like Figure~\ref{spectrum} (Right)). Then employing
Eq.~\eqref{eqW} for each value of $E(L)$ we can determine the corresponding
value of $p(L)$ and from Eq.~\eqref{luschereq} the value of $\delta(E)$. 
Near the resonance we can fit $\delta(E)$ to the relativistic Breit-Wigner
formula
\begin{equation}
\tan{\left( \delta_l -\frac{\pi}{2} \right)} = \frac{E^2-m_\phi^2}
{m_\phi \Gamma_\phi},
\end{equation}
and therefore we can determine the two parameters $m_\phi$ and $\Gamma_\phi$.
Unfortunately, the method described up to now gives poor results when
the width of the resonance is large (see the $\Delta$ resonance case, 
analyzed in~\cite{Bernard:2007cm}). 
In this case the ALCs are washed out, i.e. we cannot 
locate the presence of the resonance in the way possible using data like that 
shown in Figure~\ref{spectrum} (Right).

\subsection{Probability distribution method}
An alternative method to solve the previous problem is
based on a different way to \emph{analyze} the finite volume energy 
spectrum~\cite{Bernard:2008ax}. The basic idea is to construct the probability
distribution $W(p)$ according to the prescriptions:
\begin{enumerate}
\item Measure the two-particle spectrum $E_n(L)$ for different values of $L$
and for $n=1, \cdots, N$;
\item Determine $p_n(L)$ using Eq.~\eqref{eqW}, for each $n$ and $L$;
\item Choose a suitable momentum interval $[p_1,p_2]$ and introduce an 
equal-size momentum bin with length $\Delta{p}$;
\item Count how many eigenvalues $p_n(L)$ are contained in each bin;
\item Normalize this distribution in the interval $[p_1,p_2]$.
\end{enumerate}
It is possible to show that the probability distribution $W(p)$ is given
by $W(p)=c \sum_{n=1}^N \left[ p^\prime_n(L) \right]^{-1}$ and differentiating 
Eq.~\eqref {luschereq} with respect to $L$, it turns out:
\begin{equation}
W(p)=\frac{c}{p} \sum_{n=1}^N \left[L_n(p)+ \frac{2 \pi \delta^\prime(p)}
{\phi^\prime(q_n(p))} \right], \qquad 
\mbox{[$c$ is a normalization constant]}.
\label{probdistr}
\end{equation}
Let us introduce $W_0(p)$, determined by  Eq.~\eqref{probdistr} with 
$\delta(p)=0$ and $L_n(p)$ corresponding to the free energy levels.
The authors of~\cite{Bernard:2008ax} showed that in order to subtract the
background (free $\pi$ particles) it is convenient to consider the
subtracted probability distribution $\tilde{W}(p)=W(p)-W_0(p)$.
In the infinite volume limit, this last quantity is determined by 
$\delta(p)$ alone and close to the resonance, assuming a smooth dependence on
$p$ for the other quantities, it follows a Breit-Wigner shape with the 
\emph{same width}.
The main task of our work is to test this method on an effective field theory
where a resonance emerges and then exploit the qualitative and 
quantitative knowledge in the real case of $QCD$.

\section{A simple model of scalar meson decay}
We examine a non-linear sigma model, characterized by a field 
$\Sigma \in SU(2)$, coupled to a scalar field $\phi$ by a three-point
interaction. The action is given by
\begin{equation}
S = \int d^4x \left\{ \beta \mbox{Tr} \left( \partial_\mu \Sigma_x 
\partial_\mu \Sigma_x ^\dagger \right) 
- m_\pi^2  \beta \mbox{Tr}   \left( \Sigma_x+\Sigma_x^\dagger \right) 
+ \onehalf \phi_x (-\partial_\mu \partial_\mu+m_\phi^2)\phi_x +
\lambda\, \phi_x \, \beta \ \mbox{Tr} \left( \partial_\mu \Sigma_x 
\partial_\mu \Sigma^\dagger_x +\nu \right)   \right\} \ ,
\end{equation}
where the relation between the field $\Sigma$ and the three (pion) fields 
$\pi_i$ is given by $\Sigma_x=\exp{\left(i \frac{\pi^i_x \sigma_i}{f}\right)}$.
The coupling $\beta$ is related to the pion decay constant $f$ by 
$\beta=f^2/4$. 
The coefficient $\nu$ is introduced and tuned so that 
$\langle \phi \rangle = 0$.
The action is invariant under global chiral transformation  
$\Sigma \rightarrow U_R \Sigma U_L^\dagger$, where $U_L, U_R \in SU(2)$, apart
from an explicit symmetry breaking term  $m_\pi^2  \beta \mbox{Tr}   
\left( \Sigma_x+\Sigma_x^\dagger \right)$.
It is instructive to consider the action for $\beta \to \infty$, 
i.e. when the three pions do not interact with each other. In this limit, 
\begin{equation}
S=\onehalf \int d^4x \left\{ \pi^i_x (-\partial_\mu \partial_\mu+m_\pi^2)\pi^i_x + \phi_x (-\partial_\mu \partial_\mu+m_\phi^2)\phi_x + \lambda\, \phi_x \, \partial_\mu \pi^i_x  \, \partial_\mu \pi^i_x +\lambda f^2 \nu \phi_x \right\} \ ,
\end{equation}
and we can see clearly the characteristic coupling term where the derivatives
of the two pions appear.
When we discretize this theory on the lattice we note it is possible 
to update both fields $\phi$ and $\Sigma$ by means of the heatbath 
algorithm~\cite{Creutz:1980zw}. Naturally, this model closely resembles a linear
sigma model, although the $\phi\pi\pi$ interactions differ slightly. A ``bottom
up'' construction was used to ensure similar prescriptions could be followed to
couple mesons with general spin. An example of this for the $\rho$-meson is
found in~\cite{DeGrand:1990ip}. 

\section{Monte Carlo simulation}

\begin{figure}[tb]
  \psfrag{LAT}{$12^3\times 48$}
  \psfrag{BET}{$\beta=0.25$}
  \psfrag{LAM}{$\lambda=5.0$}
  \psfrag{MPI}{$m_\pi=0.410$}
  \psfrag{MPHI}{$m_\phi=9.0$}
  \psfrag{NU}{$\nu=4.54$}
  \psfrag{DATA}{data $120000$}
  \psfrag{mass}{mass}
  \psfrag{t}{t}
  \footnotesize
  \psfrag{N0}{$n^2=0$}
  \psfrag{N1}{$n^2=1$}
  \psfrag{N2}{$n^2=2$}
  \psfrag{N3}{$n^2=3$}
  \psfrag{N4}{$n^2=4$}
\includegraphics[width=0.37\textwidth,angle=-90]{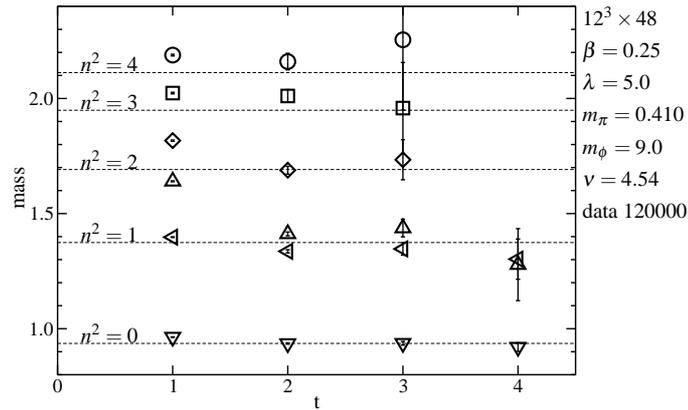}
\caption{The two-pion spectrum. The ALC is seen around the $n^2=1$ two-pion 
state. 
The dashed lines, which show the free two-pion spectrum, take in account the
discretization artefacts.}
\label{spectrmodel}
\end{figure}

In order to determine the two-particle spectrum we first introduce the
partial Fourier transform (PFT) of the field $\pi$:
\begin{equation}
\tilde{\pi}(\vec{n},t)=\frac{1}{V}\sum_x \pi(\vec{x},t) e^{i \vec{x} \vec{p}} \ , \qquad 
p_i=\frac{2 \pi}{L_i} n_i \ , \qquad n_i=0, \cdots ,L_i-1 \ .
\end{equation}
Then we study operators with zero total momentum and zero isospin:
\begin{equation}
O_{\vec{n}}(t)= \sum_{i=1}^3 \tilde{\pi}^i(\vec{n},t) 
\tilde{\pi}^i(-\vec{n},t) \ ;
\end{equation}
in particular we take in account five different operators, 
corresponding to $n^2=0,1,2,3,4$.
A sixth operator, that clearly has the correct quantum number, is the PFT 
of the field $\phi$ with $\vec{p}=0$. 
To determine the energy levels we use a method, introduced 
in~\cite{Luscher:1990ck}, based on a generalized eigenvalue problem applied 
to the correlation matrix function $C_{ij}(t)= \langle O_i O_j\rangle$, 
i.e. a matrix whose elements are all possible correlators between the six 
operators. An example of spectrum is given in Figure~\ref{spectrmodel} 
where the dashed lines represent the free two-pion spectrum.
In this plot the parameters have been chosen to show an ALC between the $\phi$
level and the $2\pi$ level for $n^2=1$. The gap between these two levels is
small compared to the difference between the others (the width of the 
resonance is small) and therefore it is simple to determine the position of 
the resonance, i.e. in this case the ALC is not washed out.

Unfortunately, it looks like it is not possible to tune the parameters of this
model to increase the width of the resonance arbitrarily.
Two different 
problems arise, depending on the value of $\beta$. If 
$\beta < 0.25$, a saturation effect appears so even if we increase the 
value of $\lambda$, the width of the resonance does not increase. 
On the other hand if $\beta$ is big, a vacuum instability effect 
prevents simulations beyond a $\lambda_{max}$ value.
At the moment, we are trying to understand the origin
of these problems in more detail through perturbative calculations.

\section{Conclusion and outlook}
The problem of studying the width of a resonance on the lattice 
has attracted much attention recently. The main tool to get results
is based  on the study of the volume dependence of the two-particle states 
spectrum. A new method to analyze the data, based on the concept of 
probability distribution, has been proposed recently.
In this work we have started to study a lattice field model where the 
new method will be tested. This model is important not only from a qualitative
point of view, but it will be quantitatively important: we can tune the 
parameters of models like this to obtain correlation functions with the same 
precision of a QCD Monte Carlo study and therefore predict the precision 
with which we can obtain the QCD resonance parameters. Investigations of the
scalar glueball following a prescription like this have begun. 


\begin{theacknowledgments}
This work is supported by Science Foundation Ireland under grant number
07/RFP/PHYF168.
\end{theacknowledgments}


\end{document}